\documentclass[12pt]{imsart}

\RequirePackage[OT1]{fontenc}
\RequirePackage{amsthm,amsmath,natbib,amssymb}
\RequirePackage{hypernat}
\usepackage{verbatim}

\startlocaldefs
\numberwithin{equation}{section}
\theoremstyle{plain}

\theoremstyle{remark}

\endlocaldefs

\begin{document}

\begin{frontmatter}

\title{Efficient Estimation For The Joint Model of Survival and Longitudinal Data}
\vspace{.25cm}
\runtitle{Efficient Estimation For The Joint Model of Survival and Longitudinal Data}

\centerline{ Khandoker Akib Mohammad$^\ast$, Yuichi Hirose, Budhi Surya and Yuan Yao}
\vspace{.15cm}
\centerline{\it Victoria University of Wellington}
\vspace{.25cm}






\begin{abstract}
In survival studies it is important to record the values of key longitudinal covariates until the occurrence of event of a subject. For this reason, it is essential to study the association between longitudinal and time-to-event outcomes using the joint model. In this paper, profile likelihood approach has been used to estimate the cumulative hazard and regression parameters from the joint model of survival and longitudinal data. Moreover, we show the asymptotic normality of the profile likelihood estimator via asymptotic expansion of the profile likelihood and obtain the explicit form of the variance estimator. The estimators of the regression parameters from the joint model are shown to be semi-parametric efficient. The numerical result of our proposed method is shown by the ‘aids’ data (from JM R-package) where we have computed the standard errors of the estimated parameters from the efficient score function and efficient information matrix.

\end{abstract}






\end{frontmatter}

\section{Introduction} 
In survival analysis, joint model of repeated measurements and time-to-event data has attracted attention in the last decade. In survival analysis, many clinical studies often deal with two types of outcome such as longitudinal response measurements and the event time of interest (e.g. death, development of disease etc.) [Rizopoulos, 2010]. In classical survival analysis, the longitudinal outcome and the event outcome can be modelled separately by using mixed effect model and Cox PH model respectively. However, to see the effect of longitudinal outcome as a time-dependent covariate on event outcome, Faucett and Thomas (1996) and Wulfsohn and Tsiatis (1997) proposed the joint models of longitudinal and survival outcomes where the longitudinal outcome and the event outcome were modelled jointly rather than separately. In clinical settings, the joint model of longitudinal and survival data has been widely used for various types of medical studies (Brown and Ibrahim, 2003; Ratcliffe, Guo and Have, 2004; Heish, Tseng and Wang, 2006; Rizopoulos, 2010; Hanson, Branscum and Johnson, 2011; Viviani, Alfo and Rizopoulos, 2014)

The efficiency and asymptotic distribution of semiparametric maximum likelihood estimator have been studied for the joint model by Zeng and Cai (2005). Moreover, Hanson et al. (2011), Chi and Ibrahim (2006), Brown and Ibrahim (2003), Xu and
Zeger (2001), and Wang and Taylor (2001) considered the Bayesian estimation approach of joint models using MCMC technique. Tsiatis and Davidian (2001) developed a
set of unbiased estimating equations that yields consistent and asymptotically
normal estimators where random effects have treated as nuisance parameters in their proposed conditional score approach. Though the non-parametric maximum likelihood estimation has been proposed by Wulfsohn and Tsiatis (1997) but they did not address the computation with implicit function in the profile likelihood estimation. Rizopoulos (2010) developed a R package (JM) to fit the joint model of longitudinal and survival data wher Gaussian quadrature rules have been used for the numerical integration with respect to the random effects. However, computational complexity increases with the dimension of the random-effects. For this reason, Rizopoulos (2012) proposed a pseudo-adaptive Gauss–Hermite quadrature rule which is considerably faster than the standard Gauss–Hermite rule. In JM package, Rizopoulos (2010) used the aids data where the standard errors of the estimated parameters have been calculated by using bootstrap methods.

In this paper, we profile out the baseline hazard function from the joint model and plugged the estimator in the likelihood function. However the problem is that the estimator of the baseline hazard function is an implicit function (Rizopoulos, 2012). We can solve the difficulty without differentiating the implicit function to show the asymptotic normality of the estimator (Theorem-3 and Theorem-4 in Section 3.3). This approach is alternative to the methodologies where the asymptotic normality of profile likelihood estimator has been studied (Hirose, 2011, 2016; Murphy and Vaart, 2016). By using the asymptotic expansion of the likelihood, we have found the explicit form of the efficient score function and established the asymptotic normality of the profile likelihood estimator. Hence we got the explicit form of the estimate of variance for the profile likelihood estimator. These results can be used in computation to calculate the variance of the profile likelihood estimator which is illustrated in the numerical example (Section-4). For the actual computation, we have used the same data (aids from JM R-package) and computed the standard errors of the estimated parameters  from the efficient score function and efficient information matrix.

This paper is organized as follows. A brief discussion on joint model of longitudinal and survival data has been given in Section-2. In Section-3, we describe the estimation procedure and theorems which are used to show that the profile likelihood estimators are consistent and asymptotically normal. Results obtained from the profile expansion of the joint model are shown in Section-4. This paper concludes in Section-5  with a short discussion.

\section{Joint Model of Survival and Longitudinal Data}
For each subject $i=1,\ldots,n$,  
we observe response $y_{i}(t)$,  fixed effects covariate vector $x_{i}(t)$ and random effects covariate vector $z_{i}(t)$. Now we assume a linear mixed model for the process as
\begin{eqnarray*}
	y_{i}(t) & = & m_{i}(t)+\varepsilon_{i}
\end{eqnarray*}
where $ m_{i}(t)=x'_{i}(t) \beta+ z'_{i}(t) b_{i}$ is the longidudinal process with the fixed effects $\beta$ and the random effects $b_{i}$. 
We assume the error $\varepsilon_{i}  \sim  N(0,\sigma^2 I_{n_i})$ and the random effects $b_{i}  \sim  N(0,D)$ are independent, where $I_{n_i}$ denotes the $n_i$-dimensionality identity matrix and $D$ is the covariance matrix.

In addition, for the survival data, let us assume $T^*_i$ and $C_i$ be the survival and censoring times, respectively. Moreover, $T_i$ be the observed event time which can be defined as $T_i=\min (T^*_i,C_i)$ and $\delta_i=I(T^*_i\leq C_i)$ be the censoring indicator. 
We observe $(T_i,\delta_i,w_i)$, where $w_i$ be a vector of baseline covariates. 
We assume there is no measurements after the event or the censorship: $t_{in_i} \leq T_i$.

To measure the strength of association between $m_i(t)$ and the risk of the event, we can express the relative risk models as

\begin{eqnarray}\label{Cox_model}
\lambda_i(t|b_i)=\lambda_0(t)\exp\left\{\gamma' w_i+\alpha' m_{i}(t) \right\}. 
\end{eqnarray}
where $\lambda_0(t)$ is the baseline hazard function and $\gamma$ is the vector of regression coefficients corresponding to $w_i$. Similarly $\alpha$ represents the effect of the underlying longitudinal outcome to the risk of the event.

Let $\Lambda(t)=\int_0^t \lambda_0(s)ds$ be the cumulative baseline hazard function. Then Survival function can be written as
$$S_i(t|b_i)=\exp \left(- \int_0^t \lambda_0(s) \exp\left\{\gamma' w_i+\alpha' m_{i}(s) \right\}ds\right),$$

\section{Estimation}
To construct the joint model, we assume that random effects $b_i$ underlies both the longitudinal and event outcomes (conditional independence). The joint likelihood can be written as
\begin{equation*}
L(\theta,\Lambda)=\prod_{i=1}^{n} \int_{b_i}^{}p(T_i, \delta_i, y_i, b_i;\theta, \Lambda)db_i=\prod_{i=1}^{n} \int_{b_i}^{} p(T_i , \delta_i| b_i;\theta, \Lambda)p(y_i| b_i;\theta)p(b_i;\theta)db_i
\end{equation*}
where 
\begin{eqnarray}\label{pdf_S}
p(T_i,\delta_i|b_i;\theta,\Lambda)
& = & \left[\lambda_0(T_i)\exp\left\{\gamma' w_i+\alpha' m_{i}(T_i) \right\}\right]^{\delta_i}\nonumber \\
&  & \times \exp \left(- \int_0^{T_i} \exp\left\{\gamma' w_i+\alpha' m_{i}(s) \right\}d\Lambda(s)\right)
\end{eqnarray}
\begin{eqnarray}\label{pdf_L1}
p(y_i|b_i;\theta) 
& = & (2\pi\sigma^2)^{-n_i/2} \exp\left\{-\frac{\sum_{j=1}^{n_i}[y_{i}(t_{ij})-m_{i}(t_{ij})]^2}{2\sigma^2}\right\}, 
\end{eqnarray}
and
\begin{eqnarray}\label{pdf_L2}
p(b_i|\theta)  
& = & 
(2\pi)^{-r/2} \det (D)^{-1/2} \exp\left\{-\frac{b_{i}'D^{-1}b_{i}}{2}\right\},
\end{eqnarray}
where $r$ is the dimension of $b_{i}$ and $\theta=(\alpha',\beta',\gamma',\sigma',\textrm{vec}(D)')'$ be the all parameters in the joint model except the cumulative baseline hazard function $\Lambda$. Moreover, ``$\textrm{vec}(D)$'' is the vectorization of the matix $D$.
In our joint model, $\theta$ is the main parameter of interest whereas $\Lambda$ is the nuisance parameter.

\subsection{EM Algorithm}
Let us define the complete data by $(t_i,\delta_i, y_i, b_i), ~i=1,..., n$ which includes the observed data and unobserved $b_i$ (treating $b_i$ as missing data). The choice for using EM algorithm is justified by the fact that the model depends on a latent variable, $b_i$. Moreover, the aim of 
EM algorithm is to maximize observed data likelihood from a complete data likelihood. So the complete data log-likelihood function can be written as
\begin{equation*}
\log L_c(\theta,\Lambda;b)=\sum_{i=1}^{n} \log \bigg\{ p(T_i , \delta_i| b_i;\theta, \Lambda)p(y_i| b_i;\theta)  p(b_i;\theta) \bigg\}.
\end{equation*}

Now the expected complete data log-likelihood under $p(b|T, \delta, y)$ will be
\begin{equation}\label{3.4}
\begin{aligned}
E_{b}[\log L_c(\theta,\Lambda;b)]={} & \sum_{i=1}^{n}  \int_{b_i}^{} \log \bigg\{ p(T_i , \delta_i| b_i;\theta, \Lambda)p(y_i| b_i;\theta)  p(b_i;\theta) \bigg\} p(b_i|T_i, \delta_i, y_i; \theta, \Lambda)db_i \\
={} & \sum_{i=1}^{n} \bigg\{  E_{b} \big[\log p(T_i , \delta_i| b_i;\theta, \Lambda) \big]+E_{b} [\log p(y_i| b_i;\theta)]+E_{b} [\log p(b_i;\theta)] \bigg\}
\end{aligned}
\end{equation}
where $E_{b}$ is the expectation with respect to $b$ under $p(b|T, \delta, y)$ which can be expressed as
\begin{equation*}
p(b|T, \delta, y; \theta, \Lambda)=\frac{p(T , \delta| b;\theta, \Lambda)p(y| b;\theta)p(b;\theta)}{\int_{b}^{} p(T , \delta| b;\theta, \Lambda)p(y| b;\theta)p(b;\theta)db} \propto p(T , \delta| b;\theta, \Lambda)p(y| b;\theta)p(b;\theta)
\end{equation*}

\subsection*{Baseline Hazard Estimation}
Before starting the EM algorithm, we profile out the baseline hazard function $\lambda_0(t)$ using NPMLE (non-parametric maximum likelihood estimation). Without loss of generality we assume observed times $t_i$ are ordered: $t_1< t_2 < \ldots <t_n$. We assume $\lambda_i$ be the hazard at time $T_i$ and treat the survival time as discrete.
The cumulative bseline hazard function is then
$\Lambda_0(t_i)=\sum_{j=1}^{n}\lambda_j1\{t_j\leq t_i\}$.

Now the survival part of equation (\ref{3.4}) can be written as
\begin{equation}
\sum_{i=1}^{n} E_{b} \bigg[\delta_i\big\{\log \lambda_i+\gamma' w_i+\alpha' m_{i}(t)\big\}-\exp \{\gamma' w_i+\alpha' m_{i}(t)\}\sum_{j=1}^{n}\lambda_j1\{t_j\leq t_i\}\bigg].
\end{equation}

Now from the derivative with respect to $\lambda_k$, we get
$$\hat{\lambda}_k(t)=\frac{\delta_k}{\sum_{l=1}^{n} 1\{t_k\leq t_l\}E_{b}[\exp \{\gamma' w_l+\alpha' m_l(t_k)\}]}.$$

So the estimate of the baseline cumulative hazard will be
\begin{equation}
\hat{\Lambda}(t)=\sum_{i=1}^{n}\frac{\delta_i1\{t_i\leq t\}}{\sum_{l=1}^{n} 1\{t \leq t_l\}E_{b}[\exp \{\gamma' w_l+\alpha' m_l(t)\}]}.
\end{equation}

\subsection*{\underline{The E-step}}
In the E-step, we use the current parameter estimates $\theta$ to find the expected values of $b$ under $p(b|T, \delta, y; \theta, \hat{\Lambda})$:
\begin{equation}
p(b|T, \delta, y; \theta, \hat{\Lambda})
\propto p(T , \delta| b;\theta, \hat{\Lambda})p(y| b;\theta)p(b;\theta)
\end{equation}
\subsection*{\underline{The M-step}}
By replacing $\lambda$ with $\hat{\lambda}(\theta)$, we maximize the equation (\ref{3.4})
\begin{equation}\label{3.8}
\begin{aligned}
E_{b}[\log L_c(\theta,\hat{\Lambda};b)]
={} & \sum_{i=1}^{n} \bigg\{  E_{b} \big[\log p(T_i , \delta_i| b_i;\theta, \hat{\Lambda}) \big]+E_{b} [\log p(y_i| b_i;\theta)]+E_{b} [\log p(b_i;\theta)] \bigg\}
\end{aligned}
\end{equation}
with respect to $\theta$ to obtain $\hat{\theta}$ respectively. The estimated parameters from the M-step are returned into E-step until the values of $\hat{\theta}$ converge.

\subsection{Score Functions}
An estimator of the baseline cumulative hazard function in the counting process notation (\citealp{fleming2011counting}) can be written from equation (3.6) as
\begin{equation}
\hat{\Lambda}(t)=\int_{0}^{t}\frac{\sum_{i=1}^{n} dN_i(u)}{\sum_{i=1}^{n} Y_i(u)E_{b}[\exp \{\gamma' w_i+\alpha' m_i(u)\}]},
\end{equation}
where $N_i(u)=1\{T_i \leq u,\delta=1\}$ and $Y_i(u)=1\{T_i \geq u\}.$

Let us denote $E_{F_n} f= \int fdF_n$. Then $\hat{\Lambda}(t)$ can be expressed as
\begin{equation}
\hat{\Lambda}_{\theta, F_n}(t)=\int_{0}^{t}\frac{E_{F_n} dN(u)}{E_{F_n} Y(u)E_{b}[\exp \{\gamma' w+\alpha' m(u)\}]}.
\end{equation}

Now from (3.8), the log-profile likelihood can be written as
\begin{equation}\label{3.11}
\begin{aligned}
E_{b}[\log L_c(\theta,\hat{\Lambda}_{\theta, F_n};b)]
={} & \sum_{i=1}^{n} \bigg\{  E_{b} \big[\log p(T_i , \delta_i| b_i;\theta, \hat{\Lambda}_{\theta, F_n}) \big]+E_{b} [\log p(y_i| b_i;\theta)]+E_{b} [\log p(b_i;\theta)] \bigg\}
\end{aligned}
\end{equation}
where $\log p(T_i , \delta_i| b_i;\theta, \hat{\Lambda}_{\theta, F_n})$, $\log p(y_i| b_i;\theta)$ and $\log p(b_i;\theta)$ are the log-profile likelihood functions (for one observation) for Cox PH, longitudinal and random effect components respectively. Now we can express the components as
\begin{equation}
\begin{aligned}
E_{b} \big[\log p(T_i , \delta_i| b_i;\theta, \hat{\Lambda}_{\theta, F_n}) \big]
={} & E_{b}\bigg[\delta_i\bigg\{\log \frac{E_{F_n} dN(T_i)}{E_{F_n} Y(T_i)E_{b}[\exp \{\gamma' w +\alpha' m(T_i)\}]}+\gamma' w_i+\alpha' m_i(T_i)\bigg\}\\
& -\exp \bigg(\gamma' w_i+\alpha' m_{i}(T_i) \bigg)
\int_{0}^{T_i}\frac{E_{F_n} dN(u)}{E_{F_n} Y(u)E_{b}[\exp \{\gamma' w+\alpha' m(u)\}]} \bigg],
\end{aligned}
\end{equation}

\begin{equation}
E_{b} [\log p(y_i| b_i;\theta)]=E_{b}\bigg[ -\frac{1}{2} \bigg\{ n_i  \log (2 \pi) +\log (\sigma^2)  +(y_i- x_i \beta- z_i b_i)' (\sigma^2)^{-1} (y_i- x_i \beta- z_i b_i)'  \bigg\} \bigg],
\end{equation}
and
\begin{equation}
E_{b} [\log p(b_i;\theta)]=E_{b}\bigg[ -\frac{1}{2} \bigg\{ r  \log (2 \pi) + \log |D|+ b_i'D^{-1} b_i  \bigg\} \bigg].
\end{equation}

The score functions for the profile likelihood are
\begin{equation}
S(T_i, \delta_i, y_i, b_i | \theta, F_n)=S_s(T_i, \delta_i|\theta, F_n)+S_l(y_i|\theta)+S_r(b_i|\theta),
\end{equation}
where
\begin{equation}
S_l(y_i|\theta)= \frac{\partial}{\partial \theta}E_{b} [\log p(y_i| b_i;\theta)],
\end{equation}
and 
\begin{equation}
S_r(b_i|\theta)= \frac{\partial}{\partial \theta}E_{b} [\log p(b_i;\theta)],
\end{equation}
are the score functions for longitudinal and random effect components respectively. Moreover, $S_s(T_i, \delta_i|\theta, F_n)$ is score function for the survival component which can be expressed as
\begin{equation}
\begin{aligned}
S_s(T_i, \delta_i|\theta, F_n)={} & \frac{\partial}{\partial \theta} E_{b} \big[\log p(T_i , \delta_i| b_i;\theta, \hat{\Lambda}_{\theta, F_n}) \big]\\
={} & E_{b} \left[    \delta_i\left\{ \left( \begin{array}{c} m_{i}(T_i) \\ \alpha x_i(T_i) \\ w_i  \end{array}\right)- \frac{L_{1,\theta,F_n}(T_i)}{L_{0,\theta,F_n}(T_i)} \right\} \right]\\
& - E_{b} \left[\int_0^{T_i} \left\{\left( \begin{array}{c} m_{i}(s) \\ \alpha x_i(s) \\ w_i  \end{array}\right) - \frac{L_{1,\theta,F_n}(s)}{L_{0,\theta,F_n}(s)} \right\}\exp\left\{\gamma' w_i+\alpha' m_{i}(s)\right\}d\hat{\Lambda}_{\theta,F_n}(s) \right].
\end{aligned}
\end{equation}
where 
\begin{eqnarray}\label{L}
L_{0,\theta,F_n}(u) & = &  E_{F_n}\left[ Y(u)  E_{b} \exp\left\{\gamma' w+ \alpha' m(u) \right\}\right],\nonumber \\
L_{1,\theta,F_n}(u) & = &  E_{F_n} \left[ Y(u)  E_{b}\left( \begin{array}{c} m(u) \\ \alpha x(u) \\ w  \end{array}\right) \exp\left\{\gamma' w+ \alpha' m(u) \right\}\right].
\end{eqnarray}

Now we will calculate the score function $B(T_i, \delta_i|\theta, F)$, which is Hadamard differentiable with respect to $F$. For an integrable function $h$ with the same domain as $F$, we can express
\begin{equation*}
\begin{aligned}
B(T_i, \delta_i|\theta, F)h={} & d_{F} E_{b} \big[\log p(T_i , \delta_i| b_i;\theta, \hat{\Lambda}_{\theta, F}) \big]h\\
={} & E_{b} \bigg\{\delta_i \bigg[\frac{E_{h} dN(T_i)}{E_{F} dN(T_i)}  -\frac{E_{h} Y(T_i)E_{b}[\exp \{\gamma' w +\alpha' m(T_i)\}]}{E_{F} Y(T_i)E_{b}[\exp \{\gamma' w +\alpha' m(T_i)\}]} \bigg] \\
& -\exp \{ {\gamma' w_i+ \alpha' m_i(T_i)} \} \int_{0}^{T_i} \frac{E_{h} dN(u)}{E_{F} Y(u)E_{b}[\exp \{\gamma' w +\alpha' m(u)\}]}\\
& +\exp \{ {\gamma' w_i+ \alpha' m_i(T_i)} \} \int_{0}^{T_i} \frac{E_{F} dN(u)E_{h} Y(u)E_{b}[\exp \{\gamma' w +\alpha' m(u)\}]}{\big(E_{F} Y(u)E_{b}[\exp \{\gamma' w +\alpha' m(u)\}]\big)^2}    \bigg\}.
\end{aligned}
\end{equation*}
where, $d_{F} \log P(T_i, \delta_i|b_i;\theta, \hat{\Lambda}_{\theta, F})$ represents the Hadamard derivative of $\log P(T_i, \delta_i|b_i;\theta, \hat{\Lambda}_{\theta, F})$ with respect to $F$ (\citealp{hirose2011efficiency}).\\  

\textbf{Theorem 1:} At the true value of $(\theta, F)$, we are going to proof the followings

1. $\hat{\Lambda}_{\theta_0, F_0}(t)=\Lambda_0(t)$, the true cumulative hazard and

2. The score function $S(T, \delta, y, b| \theta_0, F_0)$ defined in (3.15)
is the efficient score function where we drop the subscript $i$.\\

\textbf{Proof:}
Replace $F_n$ by $F_0$,we get from (3.10)
\begin{equation}
\hat{\Lambda}_{\theta_0, F_0}(t)=\int_{0}^{t}\frac{E[ dN(u)]}{E [Y(u)E_{b}\exp \{\gamma_0' w+\alpha_0' m(u)\}]}.
\end{equation}
where $E$ is the expectation with respect to the true distribution $F_0$. 
At the true value of the parameters $(\theta, F)$ we can write
\begin{equation}
E[ dN(u)]=E [Y(u)E_{b}\exp \{\gamma_0' w+\alpha_0' m(u)\}].
\end{equation}

So from this point of view, we have $\hat{\Lambda}_{\theta_0, F_0}(t)=\Lambda_0(t).$

The score function $S(T, \delta, y, b| \theta, F)=S_s(T, \delta|\theta, F)+S_l(y|\theta)+S_r(b|\theta)$ in (3.15) has three parts. We know that the longitudinal and random effect parts are parametric models those do not involve $\Lambda$, so we will work on the survival part of the score function. So the score function for the survival part at the true value of the parameters $(\theta, F)$ can be expressed as
\begin{equation}
\begin{aligned}
S^*_\theta=S_s(T, \delta|\theta_0, F_0)={} & E_{b} \left[    \delta\left\{ \left( \begin{array}{c} m(T) \\ \alpha_0 x(T) \\ w  \end{array}\right)- \frac{L_{1}(T)}{L_{0}(T)} \right\} \right]\\
& - E_{b} \left[\int_0^{T} \left\{\left( \begin{array}{c} m(s) \\ \alpha_0 x(s) \\ w  \end{array}\right) - \frac{L_{1}(s)}{L_{0}(s)} \right\}\exp\left\{\gamma_0' w+\alpha_0' m(s)\right\}d\Lambda_0(s) \right].
\end{aligned}
\end{equation}
where 
\begin{eqnarray}\label{L}
L_{0}(u) & = &  E\left[ Y(u)  E_{b} \exp\left\{\gamma_0' w+ \alpha_0' m(u) \right\}\right],\nonumber \\
L_{1}(u) & = &  E \left[ Y(u)  E_{b}\left( \begin{array}{c} m(u) \\ \alpha_0 x(u) \\ w  \end{array}\right) \exp\left\{\gamma_0' w+ \alpha_0' m(u) \right\}\right].
\end{eqnarray}
which is the efficient score function based on profile-likelihood approach. The calculation of the efficient score function based on the projection theory is given in Supplementary Materials (equation *).

\subsection{Asymptotic Normality of the MLE}

\underline{Assumptions:}

To show the asymptotic normality of the MLE and its asymptotic variance, we have to consider some assumptions. On the set of cdf functions $\digamma$, we use the sup-norm, i.e., for $F, F_0 \in \digamma$,
$$||F-F_0 ||_\infty=\sup_u|F(u)-F_0(u)|. $$
For $\rho>0$, let
$$\zeta_\rho=\{F \in \digamma:||F-F_0 ||_\infty < \rho\} .$$
The assumptions are given below

$\textbf{A1:}$ $P(T, \delta|b;\theta_0,F_0)> \delta >0$ for some positive constant $\delta > 0$ and $S(\tau)=P(T> \tau)=E[Y(\tau)]>0$.

$\textbf{A2:}$ The range of $x$, $z$ and $w$ are bounded and $\theta=(\alpha,\beta,\gamma)$ is in the compact set $\Theta$ which follows $||x|| \leq M$, $||z|| \leq M$, $||w|| \leq M$ and $||\beta|| \leq M$, $||\alpha|| \leq M$, $||\gamma|| \leq M$ for some $0<M< \infty$.

$\textbf{A3:}$ The empirical cdf $F_n$ is $\sqrt{n}$ consistent i.e. $\sqrt{n} |F_n-F_0|=O_p(1)$.

$\textbf{A4:}$ The efficient information matrix $I_s^*=E[S^*_{\theta}S^{*'}_{\theta}]$ is invertible.\\

\textbf{Theorem-2}: If the assumptions (A1-A4) hold, then

1. $\hat{\theta}_n \stackrel{P}{\rightarrow} \theta_0$ as $n \rightarrow \infty$~~and~~
2. $\hat{\Lambda}_{\hat{\theta}_n, F_n}-\Lambda_0=o_p(1)$.

The proof of Theorem-2 is given in Supplementary Materials.\\

\textbf{Theorem-3:} The score functions $S_s(T, \delta|\theta, F)$ and $B(T, \delta|\theta, F)$ are defined previously. Suppose for $(A1)$-$(A4)$, $\hat{\theta}_n \stackrel{P}{\rightarrow} \theta_0$ and $F_n \stackrel{P}{\rightarrow} F_0$ as $n \rightarrow \infty$, then we have
\begin{equation*}
E\bigg[\sqrt{n}\bigg\{S_s(T, \delta|\hat{\theta}_n,F_0)-S_s(T, \delta|\theta_0,F_0) \bigg\}\bigg]=-E\bigg[S_s(T, \delta |\theta_0,F_0)S'_s(T, \delta |\theta_0,F_0)\bigg] \bigg\{\sqrt{n}(\hat{\theta}_n- \theta_0) \bigg\}+ o_p(1),
\end{equation*}
and 
\begin{equation*}
\begin{aligned}
E\bigg[\sqrt{n}\bigg\{S_s(T, \delta|\hat{\theta}_n,F_n)-S_s(T, \delta|\hat{\theta}_n,F_0) \bigg\}\bigg]
={} & -E\bigg[S_s(T, \delta|\theta_0,F_0)B(T, \delta|\theta_0,F_0)\bigg] \bigg\{\sqrt{n}(F_n- F_0) \bigg\}\\
& +o_p \big(1+\sqrt{n}(\hat{\theta}_n- \theta_0)\big).
\end{aligned}
\end{equation*}
\textbf{Remark}: The results are obtained without assuming the derivative of the score functions $\frac{\partial}{\partial \theta}S_s(T, \delta|\theta,F)$ and $d_F B(T, \delta|\theta,F)$ exist. This result give us asymptotic expansion of profile likelihood without differentiating the score function that involve implicit function.

The proof of Theorem-3 is given in Supplementary Materials.\\

\textbf{Theorem-4:} If the assumptions $\{A1, A2, A3, A4\}$ are satisfied, then a consistent estimator $\hat{\theta}_n$ to the estimating equation
\begin{equation*}
\sum_{i=1}^{n}S_s(T_i, \delta_i|\hat{\theta}_n, F_n)=0,
\end{equation*}
is an asymptotically linear estimator for $\theta_0$ \citep{Hirose} with the efficient influence function $(I_s^*)^{-1}S_s(T, \delta|\theta_0, F_0)$, so that
$$\sqrt{n} (\hat{\theta}_n-\theta_0)=\frac{1}{\sqrt{n}}\sum_{i=1}^{n}(I_s^*)^{-1}S_s(T_i, \delta_i|\theta_0, F_0)+o_p(1) \stackrel{D}{\longrightarrow}N\{0,(I_s^*)^{-1}\},$$
where $N\{0,(I_s^*)^{-1}\}$ is a normal distribution with mean zero and variance $(I_s^*)^{-1}$. So the estimator $\hat{\beta}_n$ is efficient.

The proof of Theorem-4 is given in Supplementary Materials.

\section{Application}

\section{Discussion}

We have proposed a profile likelihood estimation approach for the joint model and calculated the efficient score function (Theorem-1). Moreover, we showed the consistency (Theorem-2), asymptotic normality and efficiency of the profile likelihood estimator (Theorem-3 and Theorem-4).  By using the profile likelihood approach, we got the explicit form of the variance estimator for profile likelihood estimator. For the numerical results, we have used the aids data from JM R-package and calculated the standard errors of the estimated parameters from the efficient score function.

\end{document}